\documentclass[final,5p,times,twocolumn]{elsarticle}

\usepackage[colorlinks,
            linkcolor=blue,
            anchorcolor=blue,
            citecolor=blue]{hyperref}
\biboptions{sort&compress}
\usepackage{amssymb}
\usepackage{comment}
\usepackage{amsmath,amsfonts}
\usepackage{algorithmic}
\usepackage{algorithm}
\usepackage{array}
\usepackage[caption=false,font=normalsize,labelfont=sf,textfont=sf]{subfig}
\usepackage{textcomp}
\usepackage{stfloats}
\usepackage{url}
\usepackage{verbatim}
\usepackage{graphicx}
\usepackage{cite}
\usepackage{overpic} 
\usepackage{subcaption}
\usepackage{booktabs,multirow} 
\usepackage{amssymb} 
\usepackage{bbding} 
\usepackage{ulem} 

\begin{document}

\begin{frontmatter}



\title{Large coordinate kernel attention network for
lightweight image super-resolution}


\cortext[cor1]{Corresponding author.}

\author[address1,address2]{Fangwei Hao}
\ead{haofangwei@mail.nankai.edu.cn}

\author[address1,address2]{Jiesheng Wu}
\ead{jasonwu@mail.nankai.edu.cn}

\author[address1,address2]{Haotian Lu}
\ead{luhaotian1204@163.com}

\author[address1,address2]{Ji Du}
\ead{1120230244@mail.nankai.edu.cn}

\author[address1,address2]{Jing Xu \corref{cor1}}
\ead{xujing@nankai.edu.cn}

\author[address1,address2]{Xiaoxuan Xu \corref{cor1}}
\ead{xuxx@nankai.edu.cn}



\address[address1]{the College of Artificial Intelligence, Nankai University, Tianjin, 300350, China}
\address[address2]{Ocean Engineering Research Center, Nankai University, Tianjin 300350, China}

\begin{abstract}
The multi-scale receptive field and large kernel attention (LKA) module have been shown to significantly improve performance in the lightweight image super-resolution task. However, existing lightweight super-resolution (SR) methods seldom pay attention to designing lightweight yet effective building block with multi-scale receptive field for local modeling, and their LKA modules face a quadratic increase in computational and memory footprints as the convolutional kernel size increases. To address the first issue, we propose the multi-scale blueprint separable convolutions (MBSConv) as highly efficient building block with multi-scale receptive field, and it can focus on the learning for the multi-scale information which is a vital component of discriminative representation. As for the second issue, we revisit the key properties of LKA in which we find that the adjacent direct interaction of local information and long-distance dependencies is crucial to provide remarkable performance. Thus, taking this into account and in order to mitigate the complexity of LKA, we propose a large coordinate kernel attention (LCKA) module which decomposes the 2D convolutional kernels of the depth-wise convolutional layers in LKA into horizontal and vertical 1-D kernels. LCKA enables the adjacent direct interaction of local information and long-distance dependencies not only in the horizontal direction but also in the vertical. Besides, LCKA allows for the direct use of extremely large kernels in the depth-wise convolutional layers to capture more contextual information which helps to significantly improve the reconstruction performance, while incurring lower computational complexity and memory footprints. Integrating MBSConv and LCKA, we propose a large coordinate kernel attention network (LCAN) with efficient learning capability for local, multi-scale, and contextual information. Extensive experiments show that our LCAN with extremely low model complexity achieves superior performance compared to other lightweight state-of-the-art SR methods.
\end{abstract}




\end{frontmatter}

\section{Introduction}
Single image super-resolution (SR), as a fundamental task in low-level vision, aims at reconstructing a high-resolution (HR) image from the input low-resolution (LR) one. Over the past decade, it has attracted the attention of intensive researchers, and great success has been achieved in this field. Dong et al. \citep{ref1} first designed a three-layer SR convolutional neural network named SRCNN which obtained surprising reconstruction performance than the traditional interpolation method \citep{ref2}. Then, numerous CNN-based approaches were proposed to continuously improve reconstruction performance. Kim et al. \citep{ref3} designed a 20-layer network termed as VDSR which achieved higher effectiveness than SRCNN, and it showed that simply increasing the network depth can improve the reconstruction performance of one SR model. After the appearance of residual learning \citep{ref4}, Lim et al. \citep{ref5} proposed an enhanced deep residual network (EDSR) by introducing the residual mechanism into the SR model, and it achieved notable reconstruction performance but incurred massive model parameters. Later, based on residual mechanism, dense connection \citep{ref6} and channel attention mechanism \citep{ref15}, Zhang et al. \citep{ref7} proposed a residual dense network (RDN) and a residual channel attention network (RCAN) \citep{ref8}, both of which were composed of hundreds of convolutional layers. Next, Wang et al. \citep{ref73_3} proposed a multi-scale attention network (MAN) which is made up of multi-scale large kernel
attention (MLKA) and gated spatial attention unit (GSAU). However, the MLKA still faces the quadratic increase in computational and memory footprints as the convolutional kernel size increases. Although these SR methods are effective SR methods, numerous parameters and serious computational burden are encountered, resulting in difficulty in applying them to portable devices with limited computing resources. Therefore, it is crucial to design lightweight and efficient SR models for the applications on these portable devices. 

\begin{figure}[!t]
\centering
\includegraphics[width=3in]{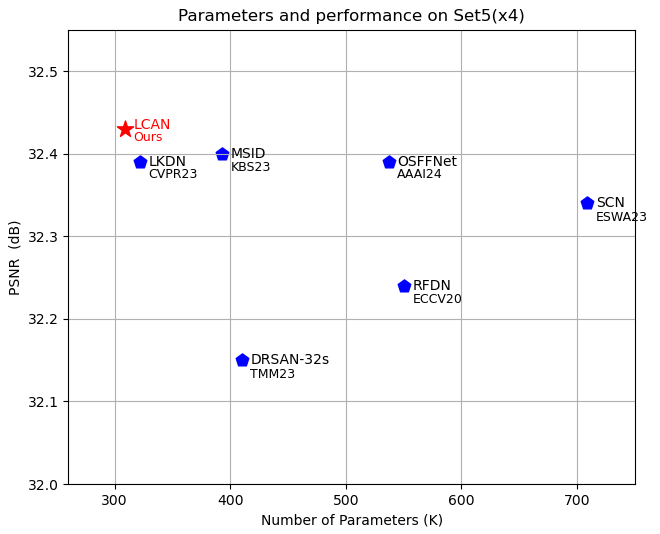}
\caption{Performance and the parameters of different methods on Set5 \citep{ref61} dataset for 4× SR. Our model achieves a better performance than previous state-of-the-art lightweight SR methods}
\label{fig_0}
\end{figure}

In terms of lightweight SR networks, by using the relatively simple recursive structure, Kim et al. \citep{ref10} presented a deep recursive convolutional network (DRCN) that used fewer parameters for large receptive field. Based on DRCN, Tai et al. \citep{ref9} further proposed a deep recursive residual network (DRRN) which also decreased the model parameters by sharing weights in the network. Despite the fact that such a parameter-sharing scheme decreases model complexity, it brings significant performance degradation. Then, some other lightweight SR endeavors \citep{ref11,ref12,ref13} are based on feature distillation strategy by channel splitting or dimension reduction. The reconstruction performance and the model efficiency of these distillation SR networks were improved partly, yet they are not efficient enough with limited performance improvement over DRRN. In addition, Sun et al. \citep{ref73_2} proposed a
spatially-adaptive feature modulation network (SAFMN) for low computational cost while achieving suboptimal performance. 

After \citep{ref14} was proposed, the attention mechanism was rapidly developed in convolutional neural networks and applied to a variety of computer vision tasks, including image classification \citep{ref15,ref16,ref17,ref18,ref19,ref20}, target detection \citep{ref21,ref22,ref23,ref24,ref25,ref26}, semantic segmentation \citep{ref27,ref28,ref29,ref30,ref31,ref32}, and image super-resolution (SR) \citep{ref33,ref34,ref35,ref36,ref37,ref38,ref38_5,ref39,ref40}.  The DRSAN \citep{ref33} was enabled by novel dynamic residual attention (DRA) and residual self-attention (RSA) module, and the LatticeNet \citep{ref37} consisted of fancy lattice blocks (LB) and contrastive loss. However, they both lacked effective contextual information, leading to limited performance in SR. Besides, SCN \citep{ref38}, which was equipped with a hierarchical self-calibration module, ignored the learning of efficient contextual information, leading to suboptimal performance with concrete model parameters. Next, Hu et al. \citep{ref39} proposed a multi-scale information distillation (MSID) network, by designing a multi-scale feature distillation (MSFD) block and constructing a scalable large kernel attention (SLKA) block via scaling attentive fields across network layers. Thanks to the multi-scale information obtained by MSFD and the large receptive field of SLKA, MSID achieved superior performance than previous lightweight SR methods.  However, single lightweight building block in MSFD cannot learn the multi-scale information, resulting in limited representation capability, and the SLKA encountered a quadratic increasing complexity as the depth-wise convolutional kernel size increases. 

Recently, Xie et al. \citep{ref40} proposed a large kernel distillation network (LKDN) by introducing an effective attention module, i.e., large kernel attention (LKA), and a new optimizer, i.e., the Adan optimizer \citep{ref41}. They utilized blueprint separable convolutions (BSConv) \citep{ref42} as the building block of LKDN, which achieved a new SOTA performance. Nevertheless, the BSConv cannot extract the important multi-scale information which is a vital component of discriminative representation, and the LKA in LKDN also encounters quadratic increasing complexity as the convolution kernel size increases. Although MSID and LKDN are both more powerful lightweight SR methods than previous lightweight ones, they seldom pay attention to designing efficient building block with multi-scale receptive field for CNNs, and their LKA modules face a quadratic increase in computational and memory footprints as the convolution kernel size increases.

To practically address these issues and further improve the reconstruction performance of one lightweight SR model, we propose a large coordinate kernel attention network (LCAN), which incorporates the designed multi-scale blueprint separable convolutions (MBSConv) and the proposed novel large coordinate kernel attention (LCKA) module. Specifically, to improve the feature extraction capability of one building convolution block, we design lightweight multi-scale blueprint separable convolutions (MBSConv) to optimize intra-kernel correlations as well as learning the multi-scale information. To take a further step, we revisit the key properties of LKA in which we find that the adjacent direct interaction of local information and long-distance dependencies is crucial to provide remarkable performance. Thus, taking this into account and in order to mitigate the complexity of LKA, we propose a large coordinate kernel attention (LCKA) module with the adjacent direct interaction of local information and long-distance dependencies in horizontal and vertical directions, respectively. The LCKA allows for the direct use of extremely large kernels in the depth-wise convolutional layers to capture more contextual information, which helps to significantly improve the reconstruction performance. Besides, LCKA incurs lower computational complexity and memory footprints than the LKA module, while keeping the advantages of LKA, including capturing local structure information, long-range dependence, and adaptability.

When equipped with the proposed MBSConv and LCKA, our
LCAN performs favorably against state-of-the-art algorithms for lightweight SR. As shown in
Figure \ref{fig_0}, LCAN with the fewest model parameters outperforms RFDN \citep{ref13}, DRSAN-32s \citep{ref33}, MSID \citep{ref39}, SCN \citep{ref38}, LKDN \citep{ref40}, and OSFFNet \citep{ref73_1}. 

Overall, our contributions are three-fold:
\begin{itemize}
\item{We propose a large coordinate kernel attention network (LCAN) which is an extremely lightweight SR model to recover a high-performance image from the LR input. Besides the effective learning capability for local, multi-scale, and contextual information, our LCAN is more lightweight than previous lightweight state-of-the-art SR networks, while achieving superior reconstruction performance.}

\item{We propose multi-scale blueprint separable convolutions (MBSConv) as a highly efficient building block with multi-scale receptive field for local modeling, and it can focus on the learning for the multi-scale information which is a vital component of discriminative representation.}

\item{We propose a novel large coordinate kernel attention (LCKA) module to capture more contextual information, while enabling the adjacent direct interaction of local information and long-distance dependencies not only in horizontal direction but also in the vertical. Compared with LKA, the LCKA incurs lower computational complexity and memory footprints, and allows for the direct use of extremely large kernels in the depth-wise convolutional layers, which boost model performance even further.}
\end{itemize}

The rest of this paper is organized as follows. Section 2 shows an overview of the related work. Section 3 presents the proposed model. Section 4 provides the empirical research results. Finally, we conclude our work in Section 5.
\section{RELATED WORK}
\subsection{Deep networks for lightweight SR}
In recent years, image super-resolution based on deep learning has made tremendous progress. Dong et al. \citep{ref1} firstly proposed the ground-breaking SR network named SRCNN, which was a three-layer convolutional neural network (CNN) and can directly model the mapping function from LR to corresponding HR. Due to the powerful representation of CNN, SRCNN achieved a significant improvement quantitatively and visually compared to early interpolation-based method \citep{ref2}. To further improve the performance, Kim et al. \citep{ref3} designed a very deep super-resolution (VDSR) network with 20 convolutional layers. Then, to utilize fewer parameters for large receptive field, Kim et al. \citep{ref10} proposed a deep recursive convolutional network (DRCN) by adopting the relatively simple recursive structure. Later, Tai et al. \citep{ref9} proposed a deep recursive residual network (DRRN), which was an improved one of DRCN. Under the same network depth, DRRN had fewer parameters but outperformed DRCN. 

To make a trade-off between performance and computation, Ahn et al. \citep{ref43} proposed CARN by combining the group convolutions and the cascading mechanism. Next, Luo et al. \citep{ref37} economically adopted two butterfly structures for combining two residual blocks, and they proposed a lightweight SR model, LatticeNet, which had relatively low computation and memory requirements. Other lightweight SR networks \citep{ref44}, \citep{ref45} were designed automatically by the neural architecture search (NAS) technology which enriched network structures. Another one effective strategy was feature distillation by channel splitting or dimension reduction. Thus, Hui et al. \citep{ref11} firstly introduced the feature distillation strategy into the SR task and proposed an information distillation network (IDN), which had the advantage of fast execution due to the comparatively few numbers of filters per layer and the use of group convolution. Besides, Hui et al. \citep{ref12} improved IDN and proposed a lightweight information multi-distillation network (IMDN) by constructing the cascaded information multi-distillation blocks (IMDB), in 
\begin{figure*}[!t]
\centering
\includegraphics[width=6in]{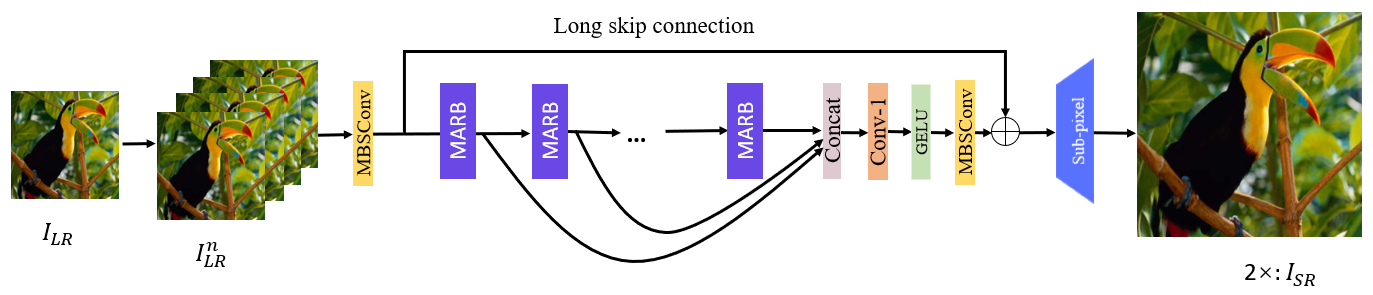}
\caption{Network architecture of our LCAN for 2× SR.}
\label{fig_1}
\end{figure*}
which the channel splitting strategy was applied multiple times and the channel-wise attention mechanism was introduced. Because of its powerful performance, the IMDN won the first place in the AIM 2019 constrained image SR challenge \citep{ref46}. Moreover, based on IMDN, Liu et al. \citep{ref13} further proposed a residual feature distillation network (RFDN) which incorporated the proposed feature distillation connection and shallow residual block. And it achieved good performance while being more lightweight. 

Later, the BSRN \citep{ref47} achieved a new state-of-the-art performance at that time. It utilized the same feature distillation structure as IMDN and introduced the blueprint separable convolutions (BSConv) \citep{ref42} to replace the standard convolution in lightweight SR,  The results in BSRN show that the BSConv is useful for reducing the parameters of the standard convolution while maintaining effectiveness. However, one BSConv lacks the ability to extract multi-scale information which is a vital component of discriminative representation. To improve the representation capacity, we design multi-scale blueprint separable convolutions (MBSConv) which incorporate the BSConv and the multi-scale structure, and we take it as the highly efficient building block of our LCAN.
\subsection{Vision attention}
Vision attention can be viewed as an adaptive reweighting according to its input feature, and its superior capability has been demonstrated in not only high-level tasks (e.g., image classification \citep{ref15,ref16,ref17,ref18,ref19,ref20}, object detection \citep{ref21,ref22,ref23,ref24,ref25,ref26}, segmentation \citep{ref27,ref28,ref29,ref30,ref31,ref32}) but also low-level tasks (e.g., image SR \citep{ref33,ref34,ref35,ref36,ref37,ref38,ref38_5,ref39,ref40}).  Since the channel attention mechanism, which was proposed in \citep{ref15}, showed its effectiveness on image classification, it and its modified ones \citep{ref48,ref49} were quickly introduced and applied to SR networks \citep{ref8,ref50,ref51}. Although the channel attention mechanism only refined the feature maps along the channel dimension, it showed significant improvements in reconstruction performance. To further improve the discriminative ability for different spatial locations, \citep{ref52} and \citep{ref53} incorporated channel-wise attention with spatial attention to force model to automatically learn the multi-level feature maps in global and local manners. \citep{ref54} and \citep{ref55} respectively proposed one unified attention module by integrating the channel-wise and spatial attention, and the generated three-dimension attention matrix was used to recalibrate feature maps in pixel-level. Besides, \citep{ref33} also adopted such joint attention, including channel-wise attention and spatial attention, to exploit discriminative representation of residual features. 

After the visual attention network (VAN) \citep{ref56} with large kernel attention (LKA) module was proposed and showed its effectiveness on various tasks, Xie et al. \citep{ref40} combined the large kernel attention (LKA) and the Adan optimizer \citep{ref41} to further propose a large kernel distillation network (LKDN), which pushed the model performance of lightweight SR to a new state-of-the-art. Although LKA can provide remarkable performance improvement, it encountered a quadratic increase in computational and memory footprints as the convolution kernel size increases. Recently, Lau et al. \citep{ref20} proposed a family of large separable kernel attention (LSKA) modules for object recognition, object detection, semantic segmentation, and robustness tests, yet no work was proposed to investigate the effect of LSKA on low-level visual tasks (e.g., image SR). By decomposing the 2D convolutional kernel of the depth-wise convolutional layers into cascaded horizontal and vertical  kernels, the LSKA incurred lower computational complexity and memory footprints than LKA, while providing similar performance. However, such decomposition ignores adjacent direct interaction of local information and long-distance dependencies in respective directions, leading to limited performance. Inspired by these works \citep{ref56},\citep{ref41},\citep{ref20}, we propose a large coordinate kernel attention (LCKA) module with adjacent direct interaction of local information and long-distance dependencies in horizontal and vertical directions for the SR task.
\section{Proposed Method}
In this section, we firstly introduce the overall network architecture of LCAN, then we give a detailed introduction to the designed multi-scale blueprint separable convolutions (MBSConv), followed by the details of the proposed novel large coordinate kernel attention (LCKA) module. Next, we introduce the proposed multi-scale attention residual block (MARB) in detail. Finally, we show the details of the loss function.

\subsection{Network Architecture}
As shown in Fig. \ref{fig_1}, similar to lightweight SR networks \citep{ref40, ref47}, our LCAN mainly consists of four parts: shallow convolution operation for shallow feature extraction, multi-scale attention residual blocks (MARBs) for deep feature extraction, the feature fusion part, and the reconstruction block. Suppose that $I_{LR}$ denotes the input LR image, and $I_{SR}$ denotes the output of our LCAN. We firstly replicate the input image $I_{LR}$ n times and concatenate the replicated images along the channel dimension to get $I_{LR}^n$ as the input of the network. Then, we use only one MBSConv, which is composed of a 1×1 convolutional layer and a multi-scale depth-wise convolution layer, to extract the multi-scale shallow feature 
\begin{equation}F_{0}=H_{s}(I_{LR}^{n}),\end{equation}
where $H_{s}(\cdot)$ is the multi-scale blueprint separation convolutions (MBSConv) operation. 
Then, the obtained $F_{0}$ is refined gradually by M stacked MARBs, each of which processes its input feature and outputs the refined one. The output refined feature of the last MARB denotes the acquired deep feature in the part of deep feature extraction, so the process of deep feature extraction can be formulated as
\begin{equation}F_{k}=H_{k}(F_{k-1}),k=1,\ldots,M,\end{equation}
where $H_{k}(\cdot)$, $F_{k-1}$, and $F_{k}$ denotes the k-th MARB operation, its input feature, and the output refined feature, respectively. After gradually refined by M MARBs, M refined features ($F_{1},\ldots,F_{M}$) from M MARBs are concatenated along the channel dimension, and one 1×1  convolution layer is used to fuse the concatenated feature maps, followed by a MBSConv for smoothing. We can formulate the process as 
\begin{equation}F_{fused}=H_{fusion}(\mathrm{Concat}(F_{1},\ldots,F_{M})),\end{equation}
where $\mathrm{Concat}(\cdot)$ denotes the concatenation operation along the channel dimension, $H_{fusion}(\cdot)$ represents the feature fusion function which is made up of a 1×1 convolution layer, a GELU \citep{ref57} activation, and one MBSConv block, and $F_{fused}$ is the obtained fused feature. At last, a long skip connection is involved across M MARBs, and the result $F_{fused}+F_{0}$ will be further processed by the reconstruction block, which consists of a 3×3 convolution layer and a sub-pixel convolution layer \citep{ref58} at the tail. We can formulate the reconstruction block as 
\begin{equation}I_{SR}=R(F_{fused}+F_{0}),\end{equation}
where $R\left(\cdot\right)$ represents the reconstruction function of the reconstruction block, and $I_{SR}$ is the reconstruction output of the entire network. 

In order to fairly compare with other state-of-the-art SR methods including LKDN, MSID, DRSAN-32s and LatticeNet-CL, we also choose the L1 loss function for model optimization. Hence, the loss function of our LCAN is formulated as
\begin{equation}
\small
L({\large{\ominus}})=\frac1N\sum_{i=1}^N\lVert H_{\mathrm{LCAN}}\left(I_{LR}^i\right)-I_{HR}^i\rVert_1\end{equation}
Where $H_{\mathrm{LCAN}}\left(\cdot\right)$ is the function of the proposed LCAN, and ${\large{\ominus}}$ represents its learnable parameters. 
In order to make a fair and comprehensive comparison with LKDN, the network is also optimized by Adan \citep{ref41} optimization algorithm, in which the adaptive optimization, decoupling weight attenuation, and modified Nesterov impulse are combined.

\subsection{Multi-scale blueprint separable convolutions (MBSConv)}
\begin{figure}[!t]
\centering
\includegraphics[width=3in]{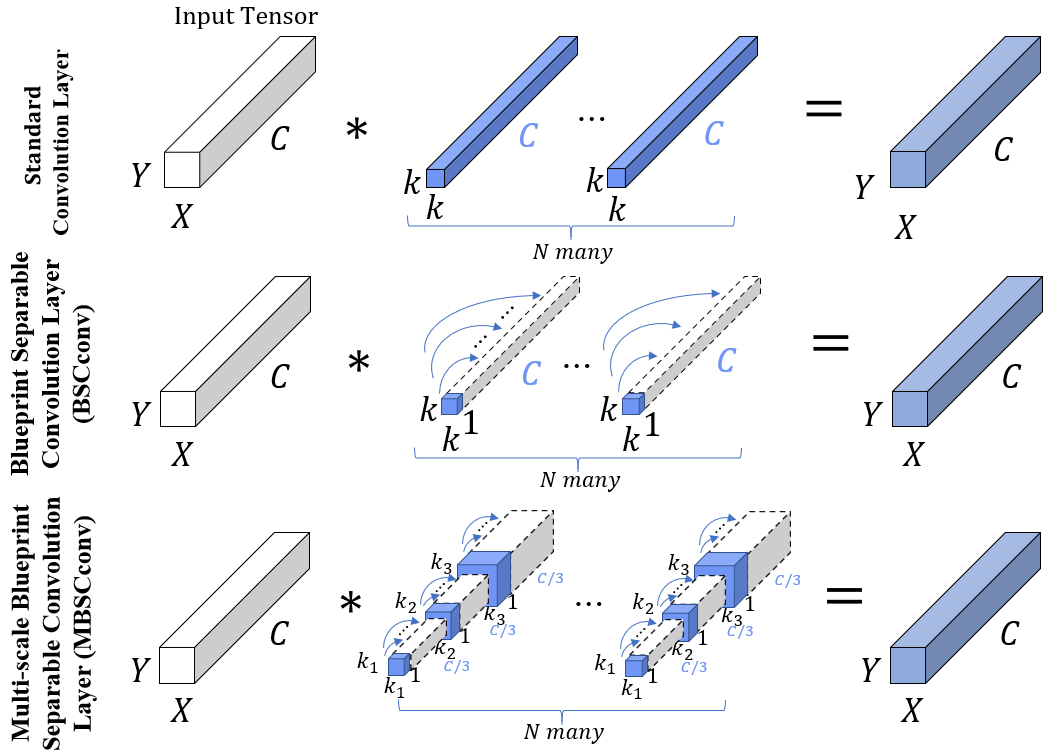}
\caption{The comparison of standard convolution layer, BSConv and the proposed MBSConv.}
\label{fig_2}
\end{figure}
Recently, improved from depth-wise separable convolution (DSConv) \citep{ref59}, a novel blueprint separation convolution (BSConv) \citep{ref42} has shown its higher learning efficiency in SR networks \citep{ref39,ref40,ref47} for optimizing intra-kernel correlations when compared to standard convolution operation. However, one BSConv lacks the capability of extracting multi-scale information which is crucial to discriminative feature. As demonstrated in \citep{ref39}, benefiting from the multi-scale receptive field, the acquired multi-scale information is an essential component for effective lightweight SR reconstruction. Nevertheless, existing lightweight SR methods seldom pay attention to designing efficient building block with multi-scale receptive field for local modeling. Inspired by these principles, we design the multi-scale blueprint separable convolutions (MBSConv), which incorporates the BSConv with the multi-scale structure, to extract discriminative feature with multi-scale information, and we take the MBSConv as the efficient building block of our LCAN.
\begin{figure*}[!t]
\centering
\includegraphics[width=4in]{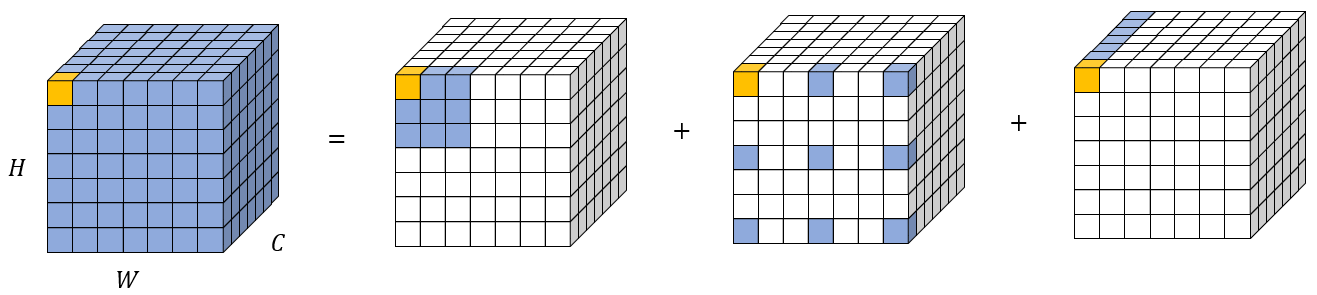}
\caption{The decomposition process of the LKA for revisiting.}
\label{fig_3}
\end{figure*}

\begin{figure*}[!t]
\centering
\includegraphics[width=6in]{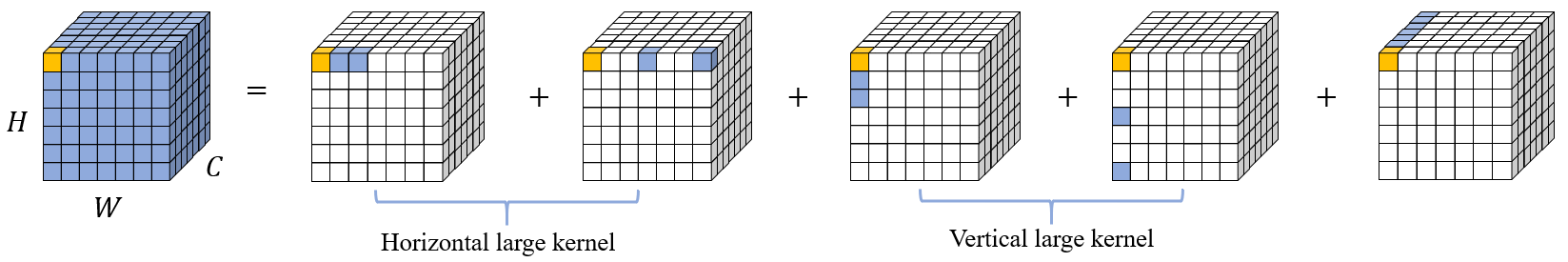}
\caption{The decomposition process of the proposed LCKA module.}
\label{fig_4}
\end{figure*}

As Fig. \ref{fig_2} shows, one MBSConv is composed of a 1×1 standard convolutional layer and a multi-scale depth-wise convolution layer with different kernel sizes. Compared to original BSConv in \citep{ref42}, our MBSConv can not only effectively formulate intra-kernel correlations and allow for a more efficient separation of regular convolutions, but also pay attention to the learning of multi-scale information for more discriminative feature. Considering the complexity in depth-wise convolution layer, we utilize the convolutional kernels with 1×1, 3×3, and 5×5. During the operation of one MBSConv, the input feature $F_{\mathrm{in}}$ is firstly refined by a 1×1 standard convolutional layer, then the refined feature maps $F_{\mathrm{refined}}$ are evenly divided into 3 parts along the channel dimension, and the divided ones are respectively processed by 1×1, 3×3, and 5×5 depth-wise convolutional kernels to generate the corresponding feature maps $F_{1\times1}$, $F_{3\times3}$, and $F_{5\times5}$. At last, $F_{1\times1}$, $F_{3\times3}$, and $F_{5\times5}$ are concatenated along the channel dimension to get feature $F_{\mathrm{MBSConv}}$ which is the output of the entire MBSConv.

\subsection{large coordinate kernel attention (LCKA) module}

\begin{figure*}[!t]
\centering
\includegraphics[width=5in]{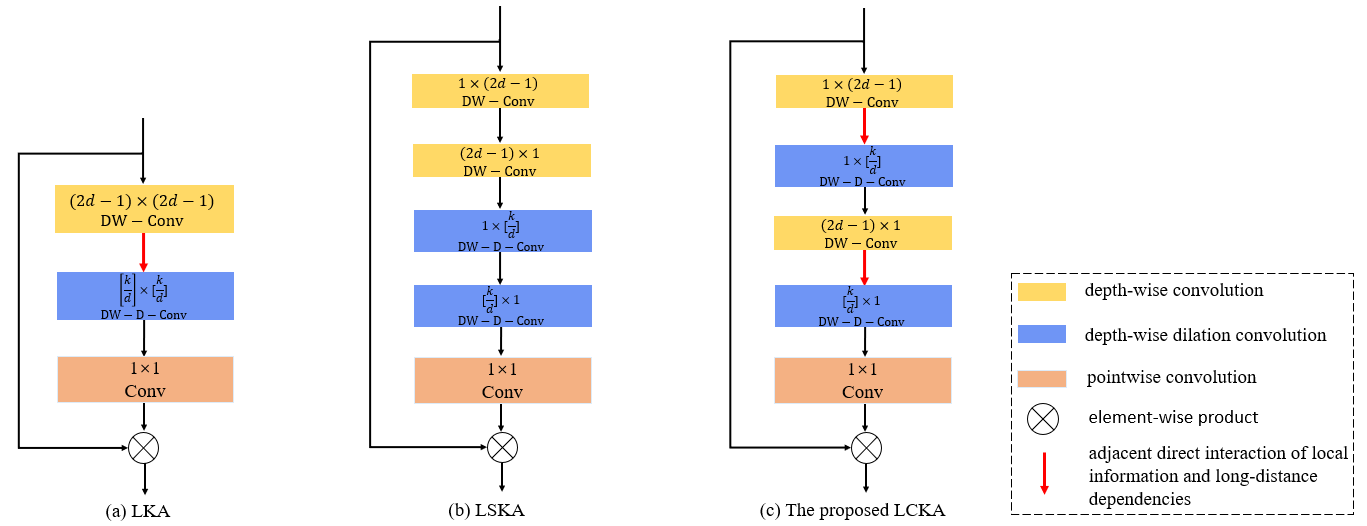}
\caption{Comparison on different designs of large kernel attention module.}
\label{fig_5}
\end{figure*}

Before introducing the LCKA in detail, we simply revisit the large kernel attention (LKA) module in \citep{ref56}. Specifically, as Fig. \ref{fig_3} shows, a 13×13 convolution can be decomposed into a 5×5 depth-wise convolution (DW-Conv5), a 5×5 depth-wise dilation convolution with dilation rate 3 (DW-D-Conv5), and a point-wise convolution (Conv1). And it can be expressed as
\begin{equation}
\small
H_{LKA13}(F_{in}^i)=H_{C1}\left(H_{DDC5}\left(H_{DC5}(F_{in}^i)\right)\right)\bigotimes F_{\mathrm{in}}^i\end{equation}
Where $H_{LKA13}(\cdot)$ is the function of LKA, $H_{DDC5}(\cdot)$, $H_{DC5}(\cdot)$, and $H_{C1}(\cdot)$ are the convolutional operations of DW-D-Conv5, DW-Conv5, and Conv1, respectively. Besides, $F_{\mathrm{in}}^i$ denotes the input feature of i-th LKA module, and $\otimes $ denotes element-wise product.  In addtion, we further revisit the key properties of LKA. As shown in Fig. \ref{fig_5}-(a) shows, within one LKA module, the spatial local convolution (depth-wise convolution) is mainly used to extract local information, then the extracted local information is immediately processed by the spatial long-range convolution (depth-wise dilation convolution) to capture long-range dependencies. Therefore, the adjacent direct interaction of local information and long-distance dependencies is a key property of LKA. In one LKA module, we can see that the complexity encounters a quadratic increase in computational and memory footprints as the kernel sizes of the depth-wise convolutions increase. 

\begin{figure*}[!t]
\centering
\includegraphics[width=6in]{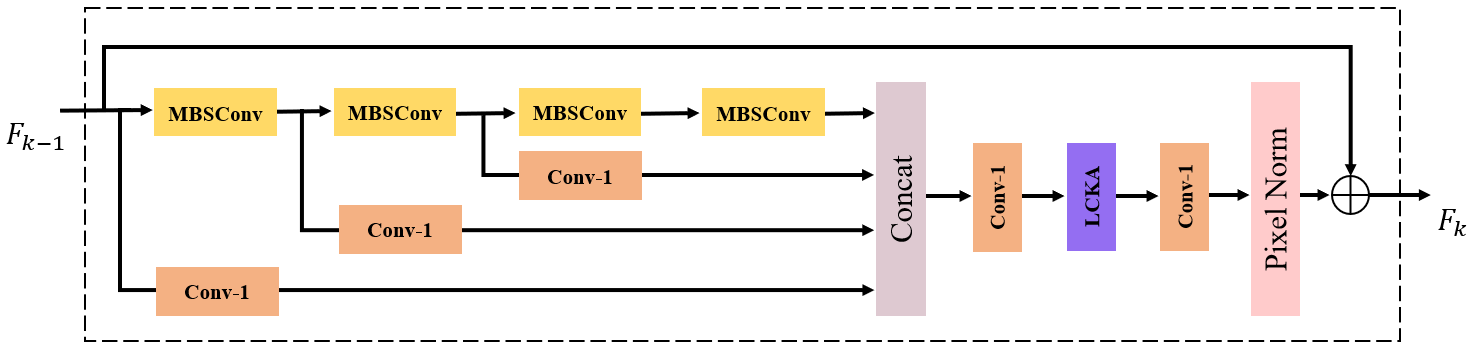}
\caption{The details of MARB.}
\label{fig_6}
\end{figure*}

 Therefore, how to keep the advantages of LKA including large receptive field and high effectiveness, while incurring lower computational complexity and memory footprints, is a key principle for building a lightweight SR module. As Fig. \ref{fig_5}-(b) shows, one method in \citep{ref20} is to decompose each depth-wise 2D convolutional kernel of $H_{DW-D-Conv5}(\cdot)$ and $H_{DW-Conv5}(\cdot)$ into staggered-connected horizontal and vertical 1-D kernels, and the decomposed attention module is denoted as LSKA. Although the LSKA incurs lower computational complexity and memory footprints than LKA, its decomposition ignores the adjacent direct interaction of local information and long-distance dependencies in respective directions, leading to limited performance. 
 
 To tackle the issue, we decompose both $H_{DW-D-Conv5}(\cdot)$ and $H_{DW-Conv5}(\cdot)$ into adjacent horizontal 1-D convolutional layers and the adjacent vertical ones, that is, the convolutional operations in each direction consist of adjacent layers of a depth-wise 1-D convolution layer and a depth-wise dilated 1-D convolution layer, the decomposition process is shown in Fig. \ref{fig_4}. Considering the obtained horizontal large kernel and vertical large kernel by decomposition, we denote the entire proposed module as large coordinate kernel attention (LCKA) module, and use it as the building attention module of our LCAN. As we can see in Fig. \ref{fig_3} and Fig. \ref{fig_4}, if under the same number of parameters, our LCKA can obtain a larger receptive field to capture more context information than LKA.
 
 In addition, as Fig. \ref{fig_5} shows, due to decomposing large convolution kernels in LKA into horizontal and vertical 1-D kernels, both LSKA and LCKA provide significant reductions in computational complexity and memory footprints, but LCKA enables the key adjacent direct interaction of local information and long-distance dependencies not only in the horizontal direction but also in the vertical. We will experimentally show that the LCKA is more effective than LSKA for the SR task.
 
\subsection{multi-scale attention residual block (MARB)}
Feature distillation strategy by channel splitting or dimension reduction in \citep{ref11,ref12,ref13}, has been demonstrated effective for the lightweight SR task. Following these works, we also use the distillation structure as the basic component of one block. Besides, in order that one block can extract more discriminative representation, we further combine it with the MBSConv and LCKA to propose a powerful multi-scale attention residual block named MARB. As we can see in Fig. \ref{fig_6}, a MARB has the advantages of feature distillation strategy, capturing multi-scale information, and efficient large coordinate kernel attention mechanism. Its high efficiency will be demonstrated in the subsequent experimental section.
\section{EXPERIMENTAL RESULTS}
In this section, the experimental setup is firstly introduced in detail, and then a series of ablation experiments on LCAN are conducted to verify the efficiency. Next, we compare our LCAN with many other state-of-the-art lightweight SR methods quantitatively
and visually. Finally, a complexity analysis of LCAN is performed.

\subsection{Settings}
Following the previous advanced works \citep{ref39,ref40,ref47}, we use the common DF2K (800 images from DIV2K \citep{ref60} and 2650 images from Flickr2K \citep{ref5}) as training dataset. After finishing training, five widely used benchmark datasets: SET5 \citep{ref61}, SET14 \citep{ref62}, BSDS100 \citep{ref63}, URBAN100 \citep{ref64} and MANGA109 \citep{ref65} are chosen as test datasets, and the output results of the SR models are converted to YCbCr space. Then, the evaluation metrics: peak signal-to-noise ratio (PSNR) and structural similarity index (SSIM) \citep{ref71} on the Y channel are adopted, and the evaluated quantitative results are applied to showing the model performance. To compare fairly with previous SR methods, we train our model on the BI training dataset which is generated by bicubic (BI) degradation with scaling factors 2×, 3× and 4×. All LR inputs have fixed patch size initially set to 48 × 48 and their mini-batch size is set to 64 during the training phase. Then, We conduct the test experiments on both BI test datasets and Real-World Photos with compression artifacts \citep{ref66,ref67}.
During training process, following \citep{ref40}, the ADAN \citep{ref41} optimizer with $\beta_{1}=0.98$, $\beta_2=0.92$ and $\beta_{13}=0.99$ is utilized to optimize our SR model from scratch, and the exponential moving average (EMA) is set to 0.999 to stabilize training. We conduct all experiments using the Pytorch \citep{ref68} framework with an NVIDIA 3090 GPU. The network is trained for $1\times10^6$ iterations, and the learning rate is set to a constant $5\times10^{-3}$.

\subsection{Ablation studies}
We conduct a series of experiments to analyze the performance of the proposed LCAN and the efficiency of its components including MBSConv and LCKA. Besides, we also conduct experiments to compare the performance of LCKA with the results of LSKA.

First, we train the model without the MBSConv, LCKA, and LSKA, on the DF2K dataset for 4× SR, and take the tested performance 32.23 dB PSNR on the Set5 (4×) dataset as the baseline. 

Then, we carry out experiments on the same datasets to train and test the model with the MBSConv only. The test result increases by 0.04 dB reaching 32.27 dB PSNR, which demonstrates that the MBSConv is effective for feature extraction and can improve the reconstruction performance by capturing discriminative feature with multi-scale information. 

Next, the model with MBSConv and LCKA is trained to verify overall performance of LCAN, and the obtained test result reaches 32.43 dB, which shows the remarkable performance of proposed LCAN for lightweight SR. 

Finally, for comparison, we carry out the same experiments on the model with MBSConv and LSKA, and corresponding test result is 32.40 dB PSNR, which is 0.03 dB lower than the result of the model with MBSConv and LCKA. The experimental results indicate that the model with MBSConv and LCKA achieves the best PSNR 32.43 dB, which demonstrates the effectiveness and superiority of proposed MBSConv and LCKA. Specifically, our MBSConv can improve the reconstruction performance by capturing discriminative feature with multi-scale information, and our LCKA can capture more contextual information by achieving extremely large receptive field with the direct interaction between local information and long-distance dependencies, which further improves model performance. All experimental results are shown in TABLE \ref{tab_1}.
\begin{table*}[htbp]
  \centering
  \caption{Investigations of MBSConv, LCKA, and LSKA on the Set5 (4×), the best results are \textbf{highlighted}.}
    \begin{tabular}{p{6em}p{3em}ccccc}
    \toprule
    \centering MBSConv & \multicolumn{1}{p{4.045em}}{\centering\XSolidBrush} & \multicolumn{1}{p{4.045em}}{\centering\Checkmark} & \multicolumn{1}{p{4.045em}}{\centering\Checkmark} & \multicolumn{1}{p{4.045em}}{\centering\Checkmark} \\
    \centering LCKA  & \multicolumn{1}{p{4.045em}}{\centering\XSolidBrush} & \multicolumn{1}{p{4.045em}}{\centering\XSolidBrush} & \multicolumn{1}{p{4.045em}}{\centering\XSolidBrush} & \multicolumn{1}{p{4.045em}}{\centering\Checkmark} \\
    \centering LSKA  & \multicolumn{1}{p{4.045em}}{\centering\XSolidBrush} & \multicolumn{1}{p{4.045em}}{\centering\XSolidBrush} & \multicolumn{1}{p{4.045em}}{\centering\Checkmark} & \multicolumn{1}{p{4.045em}}{\centering\XSolidBrush} \\
     \centering PSNR (dB) & \centering 32.23 & \centering32.27 & \centering32.40 & \textbf{32.43} \\
    \bottomrule
    \end{tabular}%
  \label{tab_1}%
\end{table*}%

\subsection{Comparisons with Advanced Methods}
To further verify the efficiency of our LCAN, we quantitatively and visually compare our experimental results for upscaling factor 2×, 3×, and 4× with the results of other state-of-the-art lightweight SR methods, including SRCNN \citep{ref1}, FSRCNN \citep{ref70}, VDSR \citep{ref3}, DRCN \citep{ref10}, DRRN \citep{ref9}, LapSRN \citep{ref69}, IDN \citep{ref11}, IMDN \citep{ref12}, PAN \citep{ref55}, RFDN \citep{ref13}, BSRN \citep{ref47}, LKDN \citep{ref40}, MSID \citep{ref39}, DRSAN-32s \citep{ref33},  LatticeNet-CL \citep{ref37}, and SCN \citep{ref38}. The self-ensemble strategy is used to further enhance our LCAN, and we denote 
it as LCAN+. 

\textbf{PSNR/SSIM results.} TABLE \ref{tab_2} shows the quantitative evaluation results for 2×, 3× and 4× lightweight SR. For 2× SR, with self-ensemble strategy, our LCAN+ has the best quantitative performance with the highest PNSR and SSIM values on all datasets. Besides, our LCAN not only achieves most of the highest PSNR and SSIM values on five benchmark datasets, but also has a minimum of 292K model parameters among these methods. For 3× SR, with a minimum of 299K model parameters among them, our LCAN+ achieves the highest PNSR and SSIM values on all datasets, offering the finest quantitative performance, and our LCAN obtains most of the highest PSNR and SSIM on five benchmark datasets. For 4× SR, all the best quantitative performance, all the second PSNR results and all the second SSIM results are reached by our LCAN+, and our LCAN, respectively. Overall, it can be seen that our LCAN can achieve superiority especially for large scaling factors (e.g., 4×) compared with previous lightweight state-of-the-art SR methods. The recorded experimental results quantitatively show that our method can yield better reconstruction performance than other state-of-the-art methods. In addition, TABLE \ref{tab_3} illustrates the comparison of model complexity and performance between the heavyweight state-of-the-art SR methods and our LCAN, and they are tested for 4× SR on the benchmark datasets. As we can see, our model with at least ten times fewer parameters can achieve competitive results even compared to the heavy SR networks. Specifically, our LCAN can respectively achieve 27.80 dB and 27.7 dB on Set14 and BSD100 datasets, which are equal to the corresponding results of the heavy network EDSR. Noted that the 0.309M parameters of our LCAN are far fewer than the 43.1M parameters of EDSR. 
For Set5 and Urban100 datasets, the performance differences start to become large, reaching 0.03$\sim$0.31dB and 0.13$\sim$0.56dB, respectively. These results suggest that our LCAN has a good trade-off between model complexity and performance.
\begin{table*}[htbp]
  \footnotesize 
  \centering
  \caption{ Quantitative results of state-of-the-art lightweight SR methods on benchmark datasets. The best and second-best results are \textbf{highlighted} and \uline{underlined}, respectively.}
    \begin{tabular}{p{8em}rp{5.5em}p{5.5em}p{5.5em}p{5.5em}p{5.5em}p{5.5em}}
    \toprule
    \multirow{2}[2]{*}{Method} & \multicolumn{1}{c}{\multirow{2}[2]{*}{Scale}} & \multirow{2}[2]{*}{Params} & Set5  & Set14 & BSD100 & Urban100 & Manga109 \\
    \multicolumn{1}{c}{} &       & \multicolumn{1}{c}{} & PSNR/SSIM & PSNR/SSIM & PSNR/SSIM & PSNR/SSIM & PSNR/SSIM \\
    \midrule
    Bicubic \citep{ref2} &       & -     & 33.66/0.9299 & 30.24/0.8688 & 29.56/0.8431 & 26.88/0.8403 & 30.80/0.9339 \\
    SRCNN \citep{ref1} &       & 8K    & 36.66/0.9542 & 32.45/0.9067 & 31.36/0.8879 & 29.50/0.8946 & 35.60/0.9663 \\
    FSRCNN \citep{ref70} &       & 13K   & 37.00/0.9558 & 32.63/0.9088 & 31.53/0.8920 & 29.88/0.9020 & 36.67/0.9710 \\
    VDSR \citep{ref3} &       & 666K  & 37.53/0.9587 & 33.03/0.9124 & 31.90/0.8960 & 30.76/0.9140 & 37.22/0.9750 \\
    DRCN \citep{ref10} &       & 1774K & 37.63/0.9588 & 33.04/0.9118 & 31.85/0.8942 & 30.75/0.9133 & 37.55/0.9732 \\
    DRRN \citep{ref9} &       & 298K & 37.74/0.9591 & 33.23/0.9136 & 32.05/0.8973 & 31.23/0.9188 & 37.88/0.9749 \\
    LapSRN \citep{ref69} & \multicolumn{1}{p{4.045em}}{\ ×2} & 251K  & 37.52/0.9591 & 32.99/0.9124 & 31.80/0.8952 & 30.41/0.9103 & 37.27/0.9740 \\
    IDN \citep{ref11} &       & 553K  & 37.83/0.9600 & 33.30/0.9148 & 32.08/0.8985 & 31.27/0.9196 & 38.01/0.9749 \\
    IMDN \citep{ref12} &       & 694K  & 38.00/0.9605 & 33.63/0.9177 & 32.19/0.8996 & 32.17/0.9283 & 38.88/0.9774 \\
    PAN \citep{ref55} &       & 261K  & 38.00/0.9605 & 33.59/0.9181 & 32.18/0.8997 & 32.01/0.9273 & 38.70/0.9773 \\
    RFDN \citep{ref13} &       & 534K  & 38.05/0.9606 & 33.68/0.9184 & 32.16/0.8994 & 32.12/0.9278 & 38.88/0.9773 \\
    LKDN \citep{ref40} & \multirow{5}[1]{*}{} & 304 K & \uline{38.12}/\textbf{0.9611} & \uline{33.90}/\uline{0.9202} & 32.27/0.9010 & 32.53/0.9322 & 39.19/\uline{0.9784} \\
    MSID \citep{ref39} &       & 375 K & {38.10}/0.9609 & {33.84}/{0.9198} & {32.29}/{0.9012} & {32.57}/{0.9326} & {39.23}/{0.9783} \\
    DRSAN-32s \citep{ref33} &       & 370 K & 37.99/0.9606 & 33.57/0.9177 & 32.16/0.8999 & 32.10/0.9279 & -/- \\
    LatticeNet-CL \citep{ref37} &       & 756 K & 38.09/0.9608 & 33.70/0.9188 & 32.21/0.9000 & 32.29/0.9291 & -/- \\
    SCN \citep{ref38} &       & 688 K & 38.10/0.9608 & 33.82/0.9200 & 32.28/0.9010 & 32.57/0.9324 & 39.12/0.9776 \\
    LCAN (ours) &       & 292 K &  \uline{38.12}/\uline{0.9610} & \uline{33.90}/\uline{0.9202} & \uline{32.30}/\uline{0.9013} & \uline{32.62}/\uline{0.9331} & \uline{39.28}/{0.9783} \\
    LCAN+ (ours) &       & 292 K &  \textbf{38.14}/\textbf{0.9611} & \textbf{33.94/0.9203} & \textbf{32.31/0.9014} & \textbf{32.78/0.9342} & \textbf{39.38}/\textbf{0.9785} \\
    \midrule
    Bicubic \citep{ref2} &       & -     & 30.39/0.8682 & 27.55/0.7742 & 27.21/0.7385 & 24.46/0.7349 & 26.95/0.8556 \\
    SRCNN \citep{ref1} &       & 8K    & 32.75/0.9090 & 29.30/0.8215 & 28.41/0.7863 & 26.24/0.7989 & 30.48/0.9117 \\
    FSRCNN \citep{ref70} &       & 13K   & 33.18/0.9140 & 29.37/0.8240 & 28.53/0.7910 & 26.43/0.8080 & 31.10/0.9210 \\
    VDSR \citep{ref3} &       & 666K  & 33.66/0.9213 & 29.77/0.8314 & 28.82/0.7976 & 27.14/0.8279 & 32.01/0.9340 \\
    DRCN \citep{ref10} &       & 1774K & 33.82/0.9226 & 29.76/0.8311 & 28.80/0.7963 & 27.15/0.8276 & 32.24/0.9343 \\
    DRRN \citep{ref9} &       & 298K & 34.03/0.9244 & 29.96/0.8349 & 28.95/0.8004 & 27.53/0.8378 & 32.71/0.9379 \\
    LapSRN \citep{ref69} & \multicolumn{1}{p{4.045em}}{\ ×3} & 502K  & 33.81/0.9220 & 29.79/0.8325 & 28.82/0.7980 & 27.07/0.8275 & 32.21/0.9350 \\
    IDN \citep{ref11} &       & 553K  & 34.11/0.9253 & 29.99/0.8354 & 28.95/0.8013 & 27.42/0.8359 & 32.71/0.9381 \\
    IMDN \citep{ref12} &       & 703K  & 34.36/0.9270 & 30.32/0.8417 & 29.09/0.8046 & 28.17/0.8519 & 33.61/0.9445 \\
    PAN \citep{ref55} &       & 261K  & 34.40/0.9271 & 30.36/0.8423 & 29.11/0.8050 & 28.11/0.8511 & 33.61/0.9448 \\
    RFDN \citep{ref13} &       & 541K  & 34.41/0.9273 & 30.34/0.8420 & 29.09/0.8042 & 28.21/0.8525 & 33.67/0.9449 \\
    LKDN \citep{ref40} &       & 311K & {34.54}/{0.9285} & {30.52}/{0.8455} & {29.21}/0.8078 & 28.50/0.8601 & 34.08/0.9475 \\
    MSID \citep{ref39} &       & 383 K & {34.54}/{0.9283} & 30.51/\uline{0.8456} & \uline{29.22}/\uline{0.8083} &{28.53}/{0.8603} & {34.14}/{0.9477} \\
    DRSAN-32s \citep{ref33} &       & 410 K & 34.41/0.9272 & 30.27/0.8413 & 29.08/0.8056 & 28.19/0.8529 & -/- \\
    LatticeNet-CL \citep{ref37} &       & 765 K & 34.46/0.9275 & 30.37/0.8422 & 29.12/0.8054 & 28.23/0.8525 & -/- \\
    SCN \citep{ref38} &       & 697 K & 34.52/0.9281 & 30.44/0.8443 & 29.15/0.8070 & 28.45/0.8582 & 33.96/0.9467 \\
    LCAN (ours) &       & 299 K & \uline{34.58}/\uline{0.9286} & \uline{30.53}/\uline{0.8456} & \uline{29.22}/{0.8082} & \uline{28.57}/\uline{0.8613} & \uline{34.18}/\uline{0.9478} \\
    LCAN+ (ours) &       & 299 K & \textbf{34.63/0.9289} & \textbf{30.57/0.8461} & \textbf{29.25}/\textbf{0.8087} & \textbf{28.68/0.8628} & \textbf{34.32/0.9485} \\
    \midrule
    Bicubic \citep{ref2} &       & -     & 28.42/0.8104 & 26.00/0.7027 & 25.96/0.6675 & 23.14/0.6577 & 24.89/0.7866 \\
    SRCNN \citep{ref1} &       & 8K    & 30.48/0.8626 & 27.50/0.7513 & 26.90/0.7101 & 24.52/0.7221 & 27.58/0.8555 \\
    FSRCNN \citep{ref70} &       & 13K   & 30.72/0.8660 & 27.61/0.7550 & 26.98/0.7150 & 24.62/0.7280 & 27.90/0.8610 \\
    VDSR \citep{ref3} &       & 666K  & 31.35/0.8838 & 28.01/0.7674 & 27.29/0.7251 & 25.18/0.7524 & 28.83/0.8870 \\
    DRCN \citep{ref10} &       & 1774K & 31.53/0.8854 & 28.02/0.7670 & 27.23/0.7233 & 25.14/0.7510 & 28.93/0.8854 \\
    DRRN \citep{ref9} &       & 298K & 31.68/0.8888 & 28.21/0.7720 & 27.38/0.7284 & 25.44/0.7638 & 29.45/0.8946 \\
    LapSRN \citep{ref69} & \multicolumn{1}{p{4.045em}}{\ ×4} & 502K  & 31.54/0.8852 & 28.09/0.7700 & 27.32/0.7275 & 25.21/0.7562 & 29.09/0.8900 \\
    IDN \citep{ref11} &       & 553K  & 31.82/0.8903 & 28.25/0.7730 & 27.41/0.7297 & 25.41/0.7632 & 29.41/0.8942 \\
    IMDN \citep{ref12} &       & 715K  & 32.21/0.8948 & 28.58/0.7811 & 27.56/0.7353 & 26.04/0.7838 & 30.45/0.9075 \\
    PAN \citep{ref55} &       & 272K  & 32.13/0.8948 & 28.61/0.7822 & 27.59/0.7363 & 26.11/0.7854 & 30.51/0.9095 \\
    RFDN \citep{ref13} &       & 550K  & 32.24/0.8952 & 28.61/0.7819 & 27.57/0.7360 & 26.11/0.7858 & 30.58/0.9089 \\
    LKDN \citep{ref40} &       & 322 K & 32.39/\uline{0.8979} & {28.79}/{0.7859} & {27.69}/{0.7402} & 26.42/0.7965 & {30.97}/{0.9140} \\
    MSID \citep{ref39} &       & 393 K & {32.40}/0.8973 & {28.79}/0.7858 & {27.69}/0.7401 & {26.43}/{0.7966} & 30.94/0.9132 \\
    DRSAN-32s \citep{ref33} &       & 410 K & 32.15/0.8935 & 28.54/0.7813 & 27.54/0.7364 & 26.06/0.7858 & -/- \\
    LatticeNet-CL \citep{ref37} &       & 777 K & 32.30/0.8958 & 28.65/0.7822 & 27.59/0.7365 & 26.19/0.7855 & -/- \\
    SCN \citep{ref38} &       & 709 K & 32.34/0.8967 & 28.71/0.7834 & 27.62/0.7381 & 26.31/0.7926 & 30.76/0.9112 \\
    OSFFNet \citep{ref38} &       & 537 K & 32.39/0.8976 & 28.75/0.7852 & 27.66/0.7393 & 26.36/0.7950 & 30.84/0.9125 \\
    LCAN (ours) &       & 309 K & \uline{32.43}/\uline{0.8979} & \uline{28.80}/\uline{0.7860} & \uline{27.70}/\uline{0.7405} & \uline{26.47}/\uline{0.7980} & \uline{31.03}/\uline{0.9145} \\
    LCAN+ (ours) &       & 309 K & \textbf{32.49}/\textbf{0.8985} & \textbf{28.84/0.7867} & \textbf{27.73/0.7412} & \textbf{26.54/0.7994} & \textbf{31.17/0.9155} \\

    \bottomrule
    \end{tabular}%
  \label{tab_2}%
\end{table*}%


\begin{table}[htbp]
  \footnotesize 
  \centering
  \caption{Quantitative comparison of our LCAN with the heavy state-of-the-art SR methods on benchmark datasets for 4× SR.}
    \begin{tabular}
    {p{2.545em}p{3.045em}p{3.045em}p{3.045em}p{3.045em}p{3.545em}}
    \toprule
    Method & Param & Set5  & Set14 & BSD100 & Urban100 \\
    \midrule
    EDSR  & 43.1M & 32.46dB & 28.80dB & 27.70dB & 26.64dB \\
    RDN   & 22.3M & 32.47dB & 28.8dB & 27.72dB & 26.60dB \\
    RCAN  & 15.6M & 32.63dB & 28.87dB & 27.77dB & 26.82dB \\
    DRN   & 9.8M  & 32.74dB & 28.98dB & 27.83dB & 27.03dB \\
    ERAN  & 8.02M & 32.66dB & 28.92dB & 27.79dB & 26.86dB \\
    LCAN (ours) & 0.309M & 32.43dB & 28.80dB & 27.70dB & 26.47dB \\
    \bottomrule
    \end{tabular}%
  \label{tab_3}%
\end{table}%

\begin{figure*}
\centering
\includegraphics[width=6in]{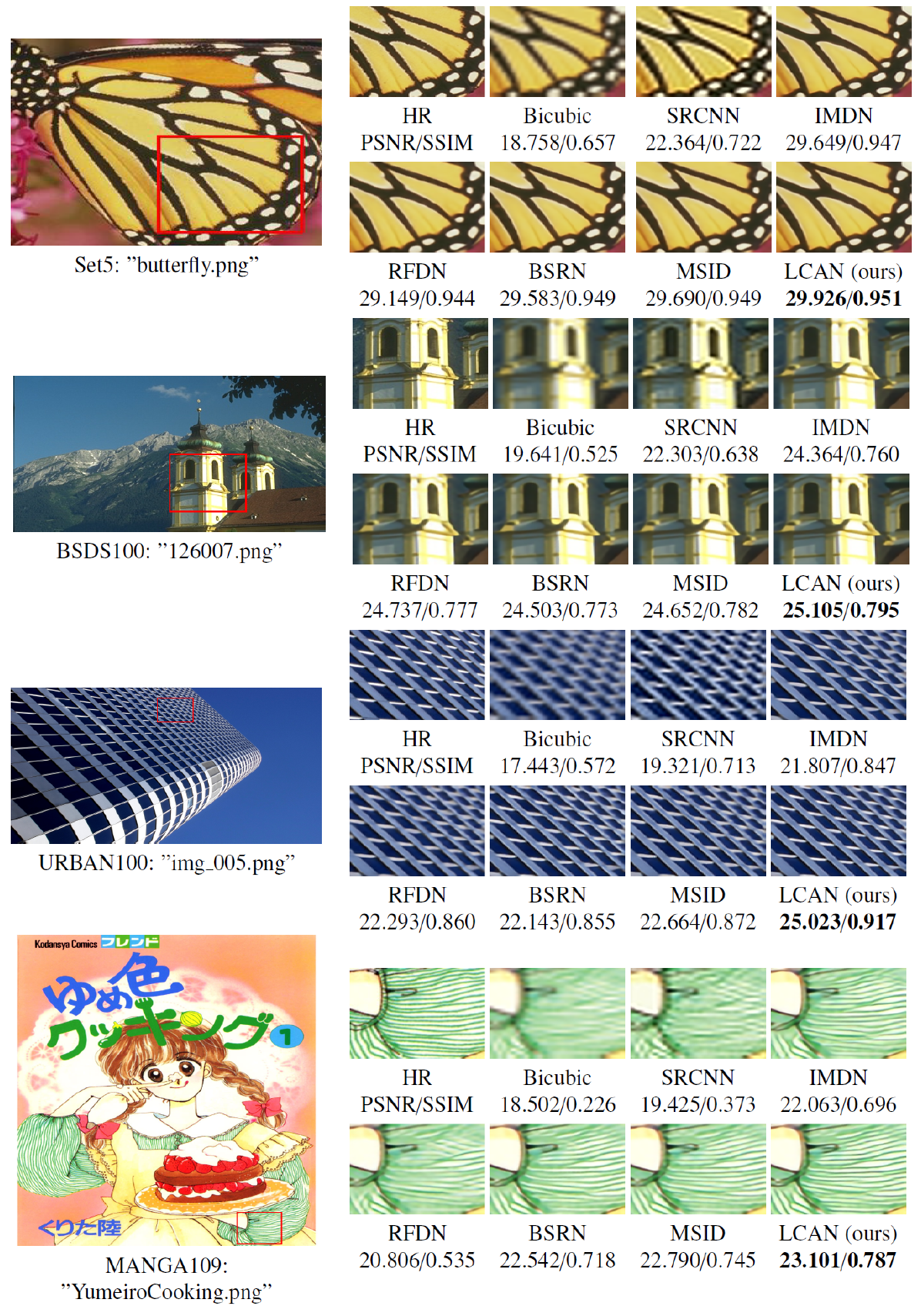}
\caption{Visual comparisons for 4× SR with the BI model on the Set5, B100, Urban100 and Manga109 datasets. The best results are highlighted. }
\label{fig_7}
\end{figure*}

\textbf{Visual results.} Fig. \ref{fig_7} presents the visual comparison of different methods on specific images of SET5, BSDS100, URBAN100 and MANGA109 datasets. As for image “butterfly.png”, a lot of sharp contour information is contained in the original HR image. From the reconstruction results of these SR methods, it can be seen that the worst visual performance is achieved by the early bicubic algorithm, while other CNN-based methods have yielded much better visual results, and the same trend goes for the quantitative results of PSNR and SSIM of these methods. For images “126007.png”, “img\_005.png” and “YumeiroCooking.png”, early bicubic algorithm still performs poorly with widespread blurring, aliasing artifacts and unstable trend, which indicates that it is an undesirable reconstruction algorithm for SR. By contrast, the visual results of other CNN-based methods (e.g., IMDN, RFDN and BSRN) have been greatly improved, they show more contour information and clearer details though they still present some distorted edges and blurring artifacts in some local areas. The quantitative results of PSNR and SSIM also show that the reconstruction results of CNN-based methods are much better than those of the bicubic interpolation algorithm. Among all these methods, our LCAN yields the best visual perception with recovering sharper detail information of contour edges, and achieves the highest PSNR and SSIM for test images, which show the effectiveness of the extracted discriminative feature with multi-scale information and the more contextual information. These results visually and quantitatively demonstrate the effectiveness and superior reconstruction performance of our method.

\subsection{Experiments on Real-World Photos}
\begin{figure}
    \centering
\includegraphics[width=3.1in]{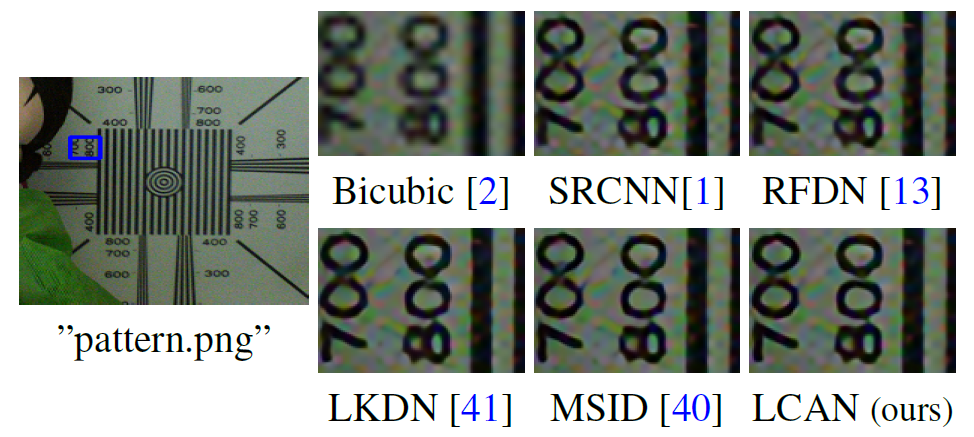}
\caption{Comparison of real image “pattern.png” for 4× SR.}
\label{fig_8}
\end{figure}

\begin{figure}
\centering
\includegraphics[width=3.1in]{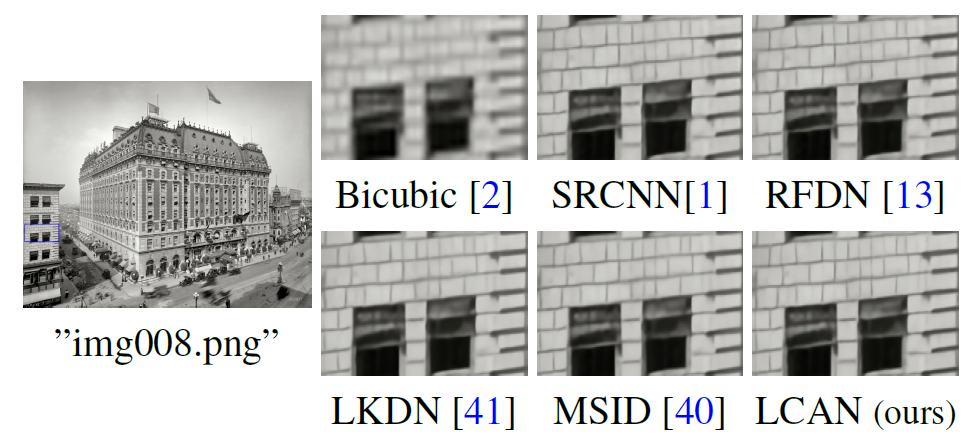}
\caption{Comparison of historical image “img008.png” for 4× SR. }
\label{fig_9}
\end{figure}

To validate the model performance in practice, we also run test experiments on real LR images that degenerate in an unknown manner. Since there are none ground-truth HR images for these real LR images, we only provide visual results for comparison. We select images “Chip” and “Historical\_006” as test images and compare our LCAN with five SR methods, including early bicubic algorithm \citep{ref2}, SRCNN \citep{ref1},  RFDN \citep{ref13}, LKDN \citep{ref40}, and MSID \citep{ref39}. As shown in Fig. \ref{fig_8} and Fig. \ref{fig_9}, our LCAN yields natural SR visual results with clearer contour information and sharper details, while early methods, i.e., bicubic and SRCNN, yield distorted and unstable results with a lot of compression artifacts, and other methods including IMDN, RFDN and LKDN produce better SR visual results than the early methods. Overall, our LCAN can yield better or comparable visual performance than other methods, not only demonstrating the effectiveness of our method but also confirming its robustness when applied to real-world images.

\begin{table*}[htbp]
\small 
  \centering
  \caption{ Computation and parameters comparison (2× Urban100),The best are \textbf{highlighted}.}
    \begin{tabular}{p{7.045em}p{5.045em}p{5.045em}p{5.045em}p{6.045em}p{7.045em}p{5.045em}}
    \toprule
    Metric & \multicolumn{1}{p{5.045em}}{IMDN \citep{ref12}} & \multicolumn{1}{p{5.045em}}{RFDN \citep{ref13}} & \multicolumn{1}{p{5.045em}}{LKDN \citep{ref40} } & \multicolumn{1}{p{7.045em}}{DRSAN-32s \citep{ref33}} & \multicolumn{1}{p{8.045em}}{LatticeNet-CL \citep{ref37}} & \multicolumn{1}{p{6.045em}}{LCAN (ours)} \\
    \midrule
    {Paras (K)} & {\quad 694} & {\quad 535} & {\quad 304} & {\quad 370} & {\quad 756} & {\textbf{\quad 291.6}} \\
    {Multi-Adds (G)} & {\quad 158.8} & {\quad 95} & {\quad 69.1} & {\quad 85.5} & {\quad 169.5} & {\textbf{\quad 68.37}} \\
    PSNR (dB) & {\quad 32.17} & {\quad 32.12} & {\quad 32.53} & {\quad 32.1} & {\quad 32.29} &  {\textbf{\quad 32.62}} \\
    
    \bottomrule
    \end{tabular}%
  \label{tab_4}%
\end{table*}%

\begin{figure}[!t]
\centering
\includegraphics[width=2.5in]{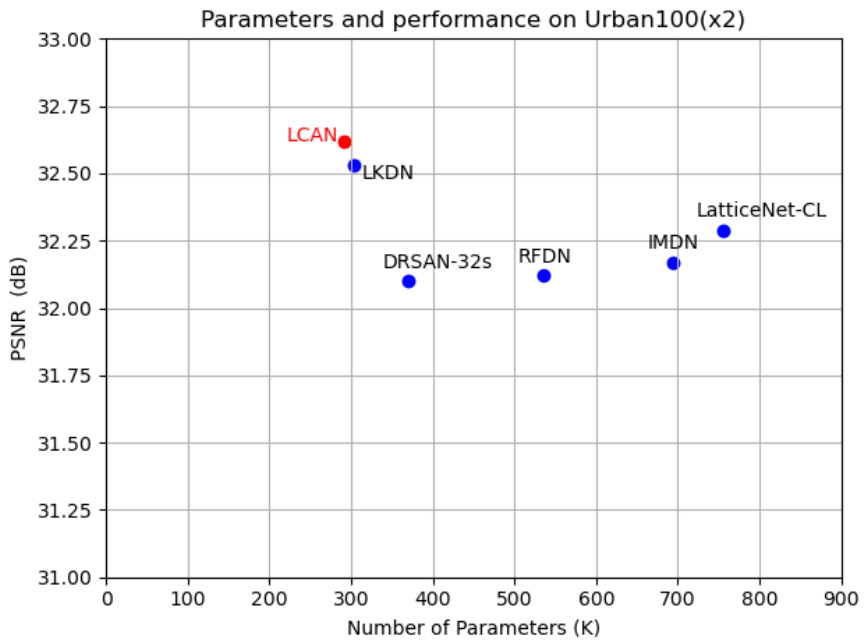}
\caption{Performance and the parameters of different methods on Urban100 (2×).}
\label{fig_10}
\end{figure}

\begin{figure}[!t]
\centering
\includegraphics[width=2.75in]{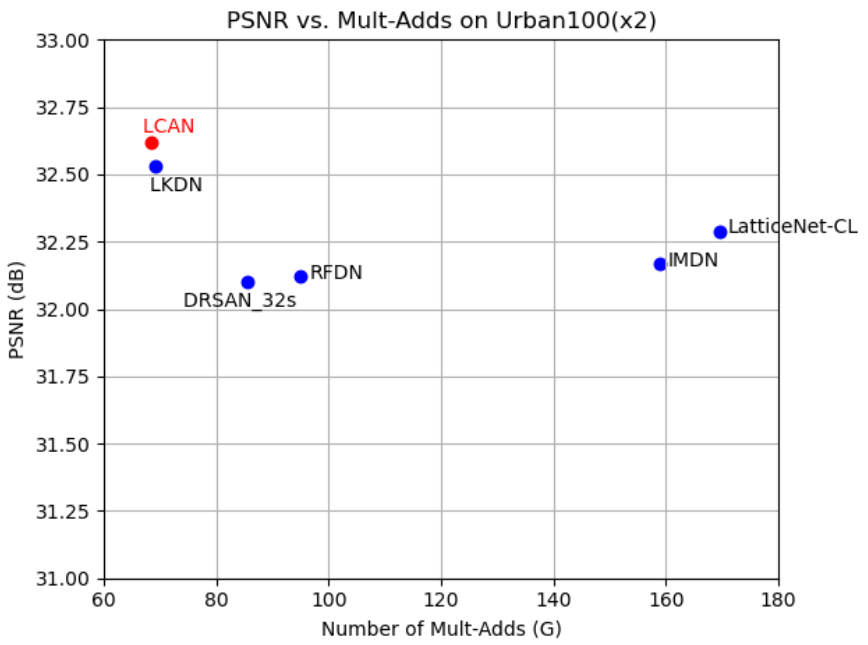}
\caption{Performance and the Multi-Adds of different methods on Urban100 (2×).}
\label{fig_11}
\end{figure}

\subsection{Model Complexity Analysis}
Now, according to the recorded results, we analyze the model complexity and reconstruction performance of different methods on Urban100 dataset for 2× SR. As widely used evaluation metrics, model parameters and multi-adds are used to quantitatively show the complexity of one SR model, and the PSNR is utilized to evaluate the reconstruction quality of visual outputs. All quantitative results of these SR methods, including LapSRN \citep{ref69}, IMDN \citep{ref12}, RFDN \citep{ref13}, LKDN \citep{ref40}, MSID \citep{ref39}, LatticeNet-CL \citep{ref37}, and our LCAN, are listed in TABLE \ref{tab_4}, and they are further illustrated in Fig. \ref{fig_10}. We can see that our LCAN achieves the highest PSNR with the fewest model parameters, which demonstrates the efficiency of our network. Noted that our LCAN with only 291.6 K parameters is the most lightweight network while achieving the highest PSNR 32.62 dB on Urban100 dataset for 2× SR.

The multi-adds, which is defined in \citep{ref43}, denotes the number of multiply accumulate operations, and it is another quantitative evaluation metric for model complexity. To fairly compare our model with other state-of-the-art SR networks, following \citep{ref30, ref40}, we assume the SR output size to 1280 × 720 to calculate model multi-adds. In order to get a more comprehensive understanding of model complexity and performance, we make a comparison of PSNR vs. Multi-Adds between our LCAN and other methods on Urban100 for 2× SR. The comparison results are showed in TABLE \ref{tab_4}, and we further visually illustrate them in Fig. \ref{fig_11}. As we can see, our LCAN has only 68.37 G multi-adds, which are 0.73 G fewer than 69.1 G multi-adds of LKDN, and it can achieve the highest PSNR 32.62 dB, outperforming LKDN by 0.09 dB and surpassing other SR methods by a large gap. The comparison results demonstrate the effectiveness and efficiency of our proposed LCAN.

\subsection{Inference time}

\begin{table*}[htbp]
\footnotesize 
  \centering
  \caption{Average inference timesin millisecond for different state-of-the-art methods on 100 SR RGB pictures with 1280 × 720 for 4× SR.}
    \begin{tabular}{p{5.045em}cccccccc}
    \toprule
    Method & \multicolumn{1}{p{5.3em}}{NGswin \citep{ref73}} & \multicolumn{1}{p{5.045em}}{\ IMDN \citep{ref12}} & \multicolumn{1}{p{5.045em}}{\ RFDN \citep{ref13}} & \multicolumn{1}{p{4.5em}}{{PAN} \citep{ref55}}  & \multicolumn{1}{p{5.155em}}{ BSRN \citep{ref47}} & \multicolumn{1}{p{5.165em}}{ LKDN \citep{ref40}} & \multicolumn{1}{p{5.045em}}{ MSID \citep{ref39}} & \multicolumn{1}{p{5.455em}}{ LCAN (ours)} \\
    \midrule
    {Time (ms)} & {157.36} & {8.93} & {7.61} & {17.26} & {17.93} & {20.97} & {18.77} & {23.95} \\
    \bottomrule
    \end{tabular}%
  \label{tab_5}%
\end{table*}%

With fewer parameters and activities, our LCAN could obtain superior performance, as shown in TABLE \ref{tab_2}. To further evaluate time efficiency, we compute the practical inference times of several state-of-the-art SR models. Using official codes of these compared methods, we compute their average inference times on one hundred 1280 × 720 super-resolved RGB images under the same experimental environment as ours. The PyTorch framework's torch.cuda.Event is used to calculate all inference times. For 4× SR, TABLE \ref{tab_5} presents the inference times for the proposed LCAN and the state-of-the-art SR methods. Theoretically, depth-wise convolution can lessen the standard convolution's computational complexity and number of parameters \citep{ref42, ref72}.

However, the depth-wise convolution's high memory access cost (MAC) to floating-point operations (FLOPs) means that the GPU's acceleration performance is currently unable to reach the theoretical value. Therefore, as TABLE \ref{tab_5} shows, our method has no obvious advantage in terms of practical inference time over other lightweight CNN-based approaches, even if our LCAN has fewer parameters and FLOPs in theory. It is worth noting that our method infers faster than transformer-based method, NGswin \citep{ref73}. Further research is required to optimize the implementation of depth-wise convolution in order to speed up its feed-forward process, which would further ease the development of lightweight networks.
\section{CONCLUSION}
We propose a large coordinate kernel attention network (LCAN) which is extremely lightweight for efficient image SR. Specifically, we propose multi-scale blueprint separable convolutions (MBSConv) to optimize intra-kernel correlations for discriminative feature with multi-scale information. In addition, we revisit the key properties of LKA in which find that the adjacent direct interaction of local information and long-distance dependencies is crucial to provide remarkable performance, thus, we propose a novel large coordinate kernel attention (LCKA) module which enables the adjacent direct interaction of local information and long-distance dependencies in horizontal and vertical directions, respectively. Besides, the LCKA allows for the direct use of extremely large kernels in the depth-wise convolutional layers to capture more contextual information, which helps to significantly improve the reconstruction performance, and it incurs lower computational complexity and memory footprints. Extensive experiments quantitatively and visually demonstrate the superiority, efficiency and robustness of our proposed LCAN for lightweight SR, on datasets generated by BI degradation and on the real-world photos.

\end{document}